\documentclass[prb,twocolumn,showpacs]{revtex4}
\usepackage{graphicx}
\usepackage{amsmath}
\usepackage{amssymb}
\usepackage[dvips]{epsfig}

\newcommand{\sss}{\scriptstyle}
\def\lsim{\
  \lower-1.2pt\vbox{\hbox{\rlap{$<$}\lower5pt\vbox{\hbox{$\sim$}}}}\ }
\def\gsim{\
  \lower-1.2pt\vbox{\hbox{\rlap{$>$}\lower5pt\vbox{\hbox{$\sim$}}}}\ }

\usepackage{enumerate, amsfonts, amsthm}

\begin{document}
\title{Possible critical regions for the ground state of a Bose gas in a spherical trap}
\author{Maksim Tomchenko} \email{mtomchenko@bitp.kiev.ua}
\affiliation{Bogolyubov Institute for Theoretical Physics, 14-b, Metrolohichna Str.,
Kiev 03680, Ukraine}

\date{\today}
\begin{abstract}
 With the help of perturbation theory, we study the ground state of a Bose gas in a spherical trap,
   using the solution in the Thomas--Fermi approximation as the zero approximation.
We have found within a certain approximation that, in some very narrow intervals of values of the magnetic field
of a trap,  the solution deviates strongly from that in
   the Thomas--Fermi approximation. If the magnetic field is equal to one
   of such critical values, the size (or even the shape) of the condensate cloud
   should significantly differ from the  Thomas--Fermi one.
\end{abstract}

\pacs{67.85.-d, 67.85.Bc}
\maketitle

\section{Introduction}
The studies of Bose-Einstein condensates in traps are intensively carried
on about two decades (see the pioneering works \cite{bec1,bec2,bec3}
and the surveys \cite{pit1999,legget2001}) and represent a
subtle complicated tool for the verification of the theory of
superfluid gas and for the solution of a number of other problems.
For interacting atoms we can approximately consider the condensate as a single
macroscopically occupied quantum state in the $\textbf{r}$-space. The most unusual are the purely
quantum effects, which have no classical analogs. A striking effect
of such a kind is the interference of two condensates
\cite{2cond}. The ground oscillatory state of the system and the
phonon excitations were registered many times and are quantum
solutions, but the shape of the condensate cloud for these states
is similar to the classical one.

It would be of interest to observe purely quantum states with
clearly nonclassical shapes of the cloud for a single condensate in
a trap. To  our knowledge, no such states have been
observed. To create them, two methods can be proposed. The first
method consists in the excitation of higher oscillatory states of the condensate with the help
of an electromagnetic miniresonator with a cylindrical or
spherical shape. The eigenmodes of such resonators have a symmetry
corresponding to the symmetry of oscillatory modes of the
condensate of the same shape. Therefore, the conservation laws
allow the condensate to absorb a quantum of electromagnetic
oscillations of the resonator and  to transit to one of the
excited oscillatory states. The experiments of such a type were
already carried out with a resonator placed in superfluid $^{4}$He
(see Ref.~\onlinecite{svh1,svh3}). In this case, a number of interesting effects
were observed, whose possible theoretical explanation was
proposed, in particular, in Ref.~\onlinecite{circ-roton}. The eigenmodes of a
disk resonator were calculated in Ref.~\onlinecite{disk-field}. The second
method consists in the search for critical points for the ground
state of a condensate, near which the solution is unstable
(nonstationary case) or ceases to be a solution (stationary case).
In vicinities of the critical points, the system is able to
spontaneously transit in the other state, in which the condensate
cloud shape can turn out quite nonclassical. In the present work,
we will seek the critical points for the stationary case. They can
exist due to the nonlinearity of the Gross--Pitaevskii (GP)
equation. As is known, the nonlinear systems are characterized by
a number of specific features such as the soliton solutions
\cite{dodd,pethick2008} and singular points in the coordinate space
\cite{blaquiere}. The value of the parameter on the boundary of
two regions corresponding to different types of singular points is
called critical. In its vicinity, the system can stepwise transit into the
other state at a small change in the parameter. For the
complicated systems, such a beautiful effect is called the
``butterfly effect'' sometimes. In what follows, we will see that
the critical points exist, apparently, also for a Bose gas in a
trap.

All values of parameters,
at which the solution differs significantly from that in the Thomas--Fermi approximation,
will be called critical values. The values of parameters,
at which the determinant of the characteristic matrix (see below) becomes zero, will be called the critical points.
The critical values generate many narrow critical regions. The critical point is located approximately at the center
of the critical region.

\section{Finding of the critical points}
For a spherical trap, the stationary GP equation takes the form
\begin{eqnarray}
E\Psi(\textbf{r})&=& -\frac{\hbar^2}{2m}\triangle \Psi + \frac{m \omega^{2}r^{2}}{2}\Psi
 + N\nu(0) |\Psi|^{2} \Psi,
     \label{1} \end{eqnarray}
where $N$ is the total number of atoms in the condensate, and
\begin{equation}
 \nu(0) = \int dV U(r)=\frac{4\pi \hbar^{2}a}{m}.
        \label{2} \end{equation}
We consider the interaction to be repulsive ($\nu(0)> 0$). Indeed, for $\nu(0)< 0$
and large $N,$ the gas collapses, and the Thomas--Fermi-type approximation is not
valid \cite{ruprecht1995,pit1999}.
It is convenient to rewrite Eq. (\ref{1}) as
\begin{eqnarray}
E\Psi(\textbf{r})&=& -\frac{\hbar^2}{2m}\frac{1}{r}\frac{\partial^{2}}{\partial r^{2}} (r\Psi) \nonumber \\
&+& \frac{\hat{L}^{2}}{2m r^{2}}\Psi + \frac{m \omega^{2}r^{2}}{2}\Psi  + N\nu(0) |\Psi|^{2} \Psi,
     \label{3} \end{eqnarray}
\begin{equation}
\hat{L}^{2} = -\hbar^{2}\left [\frac{1}{\sin^{2}{\theta}}\frac{\partial^{2}}{\partial \varphi^{2}}+
\frac{1}{\sin{\theta}}\frac{\partial}{\partial \theta} \left (\sin{\theta}\frac{\partial}{\partial \theta}\right) \right ].
        \label{4} \end{equation}
For a sufficiently large radius $R$ of the cloud,
the ground state of the gas in a trap is well described \cite{edwards1995,edwards1996,stringari1996,hau1998,pit1999}
by the formula called usually the Thomas--Fermi approximation:
\begin{equation}
\Psi^{s}_{0}(\textbf{r}) = A_{s}\sqrt{1-r^{2}/R^{2}}, \quad A_{s}=\sqrt{15/8\pi R^{3}}.
        \label{6} \end{equation}
In this case, the chemical potential \cite{legget2001} $E=E^{s}_{0}$ is
\begin{equation}
E^{s}_{0} = \frac{m\omega^{2}R^{2}}{2}= A^{2}_{s}N\nu(0).
        \label{7} \end{equation}
From whence,
\begin{equation}
R = \left (\frac{15 N\nu(0)}{4\pi m\omega^{2}}\right )^{1/5}=a_{ho}\left (\frac{15 Na}{a_{ho}}\right )^{1/5},
        \label{8} \end{equation}
where  $a_{ho} = \sqrt{\hbar/m\omega}$. Let us set
\begin{equation}
\chi(\textbf{r}) = r\Psi(\textbf{r}).
        \label{9} \end{equation}
Instead of (\ref{3}), we obtain
\begin{eqnarray}
E\chi(\textbf{r})&=& -\frac{\hbar^2}{2m}\frac{\partial^{2}}{\partial r^{2}} \chi \nonumber \\
&+& \frac{\hat{L}^{2}}{2m r^{2}}\chi + \frac{m \omega^{2}r^{2}}{2}\chi  + \frac{N\nu(0)}{r^2} |\chi|^{2} \chi.
     \label{10} \end{eqnarray}
The comparison of solution  (\ref{6})--(\ref{8}) with the numerical ones
indicates \cite{edwards1995,edwards1996,stringari1996,hau1998,pit1999},
that the approximate solution (\ref{6})--(\ref{8}) describes the system with good
accuracy, if $p\equiv R/a_{ho}=(15 Na/a_{ho})^{1/5} \gsim 4$.
Therefore, for $p \gsim 4$ it is natural to seek the exact solution of Eq. (\ref{10})  by
perturbation theory. In the first approximation, we have
\begin{equation}
\chi(\textbf{r}) \approx \chi_{0}(\textbf{r})+f(\textbf{r}), \quad E=E^{s}_{0} + \delta E,
        \label{11} \end{equation}
where $\chi_{0}(\textbf{r})=r\Psi^{s}_{0}(\textbf{r})$, and $f(\textbf{r})$ and $\delta E$ are small corrections.
According to results \cite{edwards1995,stringari1996,pit1999}
and to those obtained below, the numerical solution for $r
\rightarrow R$ deviates considerably from (\ref{6}). In addition,
the first and second derivatives of $\Psi^{s}_{0}$ (\ref{6}) with respect to
$r$  at the point $r = R$ turn to infinity. Therefore, it
is better to set $\chi_{0}(\textbf{r})$ more exactly as
\begin{equation}
\chi_{0}(\textbf{r}) =\left [ \begin{array}{ccc}
    r\Psi^{s}_{0}(r),  & \   r \leq R-\delta,   & \\
    \chi_{+}(r),  & r > R-\delta,   &
 \end{array} \right.        \label{12} \end{equation}
where  $\delta > 0$ is some small distance, and  $\chi_{+}(r)$ is the ``tail'' of
$\chi_{0}(\textbf{r})$, which is introduced formally in order to obtain a more exact description.
This tail must be sewed continuously with $r\Psi^{s}_{0}(r)$
at the point  $r = R-\delta$, have no unbounded derivatives, and have the proper asymptotics
$\exp{(-r^{2}/2a^{2}_{ho})}$ as $r \rightarrow \infty$. It is not easy to find $\chi_{+}(r)$ analytically, and we do not make it.
We consider that $\chi_{0}(r)$ is continuous on the whole semiaxis $r \in [0, \infty]$ and $\chi_{+}(r)\neq 0$.
Since $\chi_{+}(r)$ is small, we will set $\chi_{+}(r)=0$ in integrals eventually.

Let us substitute (\ref{11}) in (\ref{10})  and retain only the terms linear
in small  $f(\textbf{r})$ and $\delta E$.
We obtain
\begin{eqnarray}
&&E^{s}_{0}(\chi_{0}+f)+\delta E \chi_{0} = -\frac{\hbar^2}{2m}\frac{\partial^{2}}{\partial r^{2}} (\chi_{0}+f)
+ \frac{\hat{L}^{2}f}{2m r^{2}}\nonumber \\
&+&  \frac{m \omega^{2}r^{2}}{2}(\chi_{0}+f)  +\frac{N\nu(0)}{r^2} |\chi_{0}|^{2} (\chi_{0}+2f+f^{*}).
     \label{13} \end{eqnarray}
Since at $r \leq R-\delta$ the relation
\begin{equation}
E^{s}_{0} =  \frac{m \omega^{2}r^{2}}{2}  + \frac{N\nu(0)}{r^2} |\chi_{0}|^{2}
     \label{14} \end{equation}
holds, Eq. (\ref{13}) is reduced  in this region to
\begin{eqnarray}
\delta E \chi_{0}(r) &=& -\frac{\hbar^2}{2m}\frac{\partial^{2}}{\partial r^{2}} (\chi_{0}(r)+f(\textbf{r}))
+ \frac{\hat{L}^{2}f(\textbf{r})}{2m r^{2}} \nonumber \\
&+&    \frac{N\nu(0)}{r^2} |\chi_{0}|^{2} (f(\textbf{r})+f^{*}(\textbf{r})).
     \label{15} \end{eqnarray}
 We now expand $f(\textbf{r})$ in the full collection of eigenfunctions
of the linear problem (Eq. (\ref{10}) with $\nu(0)=0$). We collect the terms with $l=m=0$, which depend only on $r$,
in $f_{1}(r)$,  and the terms with $l\neq 0$, which depend also on $\theta$ and, possibly, on $\varphi$,
are gathered in $f_{2}(\textbf{r})$:
 \begin{equation}
f(\textbf{r}) =  \sum\limits_{nlm}c_{nlm}F_{nl}(r)Y_{lm}(\theta,\varphi)=f_{1}(r)+f_{2}(\textbf{r}).
     \label{16} \end{equation}
Then, we separate an arbitrary $(nlm)$ harmonic  from $f_{2}(\textbf{r})$. Relations (\ref{15}) and
\[\hat{L}^{2}F_{nl}(r)Y_{lm}(\theta,\varphi)=\hbar^{2}l(l+1)F_{nl}(r)Y_{lm}(\theta,\varphi),\]
\begin{eqnarray}
&&\left [ -\frac{\hbar^2}{2m}\frac{\partial^{2}}{\partial r^{2}}  + \frac{m \omega^{2}r^{2}}{2}
+ \frac{\hbar^{2}l(l+1)}{2m r^{2}}\right ]F_{nl}(r)Y_{lm}(\theta,\varphi)  \nonumber \\
&&= E^{free}_{nl}F_{nl}(r)Y_{lm}(\theta,\varphi)
     \label{17} \end{eqnarray}
yield the following equation for this harmonic (at $r \leq R-\delta$):
\begin{eqnarray}
&&c_{nlm}\left (E^{free}_{nl}-\frac{m \omega^{2}r^{2}}{2}\right )F_{nl}(r)Y_{lm}(\theta,\varphi) \nonumber \\
&&+  N\nu(0) A^{2}_{s}\left (1-r^{2}/R^{2}\right )F_{nl}(r) \nonumber \\
&&\cdot \left (c_{nlm}Y_{lm}(\theta,\varphi)+c^{*}_{nlm}Y_{lm}^{*}(\theta,\varphi)\right )=0.
     \label{18} \end{eqnarray}
If $c_{nlm}\neq 0$, then 1)
for $m\neq 0,$ the function $c_{nlm}Y_{lm}+c^{*}_{nlm}Y_{lm}^{*}$ is real, and $c_{nlm}Y_{lm}$
contains a nonzero imaginary part; it is easy to see that Eq. (\ref{18}) is not satisfied;
2) for $m= 0,$ $Y_{lm}$ is real, and the equation is satisfied for $c_{nlm}=c^{*}_{nlm}$,
 $E^{free}_{nl}=-2N\nu(0) A^{2}_{s},$ and
$m \omega^{2}/2=-2N\nu(0) A^{2}_{s}/R^{2}$, which is impossible due to the
positivity of $\nu(0)$, $E^{free}_{nl},$ and $\omega^{2}$. Therefore, the unique solution is $c_{nlm}=0.$
This implyies $f_{2}(\textbf{r})=0$ and
 \begin{equation}
f(\textbf{r}) =  f(r)=f^{*}(r), \quad \hat{L}^{2}f(\textbf{r})=0.
     \label{19} \end{equation}
In this case, Eq. (\ref{15}) is simplified:
\begin{eqnarray}
\delta E \chi_{0}(r) &=& -\frac{\hbar^2}{2m}\frac{\partial^{2}}{\partial r^{2}} (\chi_{0}(r)+f(r)) \nonumber \\
& +&    \frac{N\nu(0)}{r^2} |\chi_{0}(r)|^{2} 2f(r).
     \label{20} \end{eqnarray}
Making the changes $r=\rho a_{ho}$ and $\delta E = \varepsilon \hbar \omega/2,$ we pass to the dimensionless equation
\begin{eqnarray}
\ddot{\chi}_{0}(\rho)+\ddot{f}(\rho) +\varepsilon \chi_{0}(\rho)-
\frac{2N\nu(0)\chi_{0}^{2}(\rho) f(\rho)}{a_{ho}^{2}\rho^{2}\hbar \omega/2}=0,
     \label{21} \end{eqnarray}
where $\ddot{f}=\partial^{2}f(\rho)/\partial \rho^{2}$.  With regard for relations (\ref{12}), (\ref{6}),
$R/a_{ho}=p,$ and $2N\nu(0) A^{2}_{s}=\hbar \omega p^{2},$ Eq. (\ref{21}) takes the form
\begin{eqnarray}
 \ddot{\chi}_{0}(\rho)+\ddot{f}(\rho) +\varepsilon \chi_{0}(\rho)+2(\rho^{2}-p^{2}) f(\rho)=0.
     \label{22} \end{eqnarray}
  This equation holds for $\rho \leq p-\delta/a_{ho}$.
For $\rho > p-\delta/a_{ho},$ the more general Eq. (\ref{13}) is valid; but
since $\chi_{0}(\rho)=\chi_{+}(\rho)$ and $f(\rho)$ are small in this region,
we will solve only the simpler equation (\ref{22}), by assuming that taking the tail into account
and passing to (\ref{13})  will change the answer insignificantly.

The normalization condition is as follows:
 \begin{equation}
\int\limits_{0}^{\infty}|\Psi|^{2}(r)4\pi r^{2}dr =  4\pi\int\limits_{0}^{\infty}(\chi_{0}(r)+f(r))^{2}dr=1.
     \label{23} \end{equation}
With regard for the normalization $4\pi\int\limits_{0}^{R}\chi_{0}^{2}(r)dr=1$ for $\chi_{0}$
and the smallness of $\chi_{+}^{2}(r)$ and  $f^{2}(r)$ for $r > R$, relation (\ref{23}) yields
\begin{equation}
\int\limits_{0}^{\infty}\chi_{0}(r)f(r)dr =a_{ho}\int\limits_{0}^{\infty}\chi_{0}(\rho)f(\rho)d\rho \approx 0.
     \label{24} \end{equation}

Below, we deal with Eqs. (\ref{22}) and (\ref{24}). Let us expand
$\chi_{0}(r)$ and  $f(r)$ in the eigenfunctions of the linear
problem (\ref{17}). Since the functions  $\chi_{0}$ and $f$ depend
only on $r$, from the total collection
$\{F_{nl}(r)Y_{lm}(\theta,\varphi)\}$ it is necessary to retain only the functions
$F_{n0}(r)Y_{00}(\theta,\varphi)\equiv \Psi_{n}(r)$ in the
expansion:
 \begin{equation}
\chi_{0}(r) =  \sum\limits_{n}d_{n}\Psi_{n}(r), \quad f(r) =  \sum\limits_{n}b_{n}\Psi_{n}(r).
     \label{25} \end{equation}
We now note that, for $l=m=0,$ Eq. (\ref{17}) coincides with the equation for a one-dimensional oscillator
\begin{eqnarray}
 -\frac{\hbar^2}{2m}\frac{\partial^{2}}{\partial r^{2}}\Psi_{n}(r)  + \frac{m \omega^{2}r^{2}}{2}
\Psi_{n}(r)= E^{free}_{n}\Psi_{n}(r).
     \label{26} \end{eqnarray}
However, instead of the usual normalization $\int\limits_{-\infty}^{\infty}\Psi_{n}^{2}(x)dx=1$,
in our case, the normalization looks as
\begin{equation}
4\pi\int\limits_{0}^{\infty}\Psi_{n}^{2}(r)dr=1.
     \label{27} \end{equation}
In addition, we have $\Psi(\textbf{r})=\chi(\textbf{r})/r$. As $r \rightarrow 0,$ the wave function $\Psi(\textbf{r})$
should remain finite. This holds if we expand  the function
 $\chi(\textbf{r})=\chi_{0}(\textbf{r}) +f(r)$ in the series in $\Psi_{n}(r)$
 with only odd $n$ ($n=2j+1$, $j=0, 1, 2, \ldots$). Thus, the basis functions in (\ref{25}) are the known solutions for a
one-dimensional oscillator
\begin{equation}
\Psi_{2j+1}(r=a_{ho}\rho)=C_{2j+1}H_{2j+1}(\rho)e^{-\rho^{2}/2},
     \label{28} \end{equation}
with normalization (\ref{27}) and the energy levels
\begin{equation}
E^{free}_{2j+1}=\hbar \omega (2j+3/2).
     \label{29} \end{equation}
 In the numerical analysis, it is convenient to obtain the Hermite polynomials
\begin{equation}
H_{2j+1}(\rho)=\sum\limits_{k=0}^{j}a_{2k+1}\rho^{2k+1},
     \label{30} \end{equation}
making  use the recurrence relation \cite{vac}
 \begin{eqnarray}
a_{2k+3}&=&a_{2k+1}\frac{4k+3-2E^{free}_{2j+1}/\hbar \omega}{(2k+3)(2k+2)} \nonumber \\
&=& a_{2k+1}\frac{k-j}{(k+3/2)(k+1)}.
     \label{31} \end{eqnarray}

One can also find all solutions  of Eq. (\ref{17}) (including $l,m \neq 0$)
and verify that the subclass of solutions with $l=m=0$
corresponds to formulas (\ref{27})--(\ref{31}).

We now find the solutions of Eqs. (\ref{22}) and (\ref{24}). Since we neglect the small values of
$\chi_{0}(\rho)$ and  $f(\rho)$ for $\rho > p-\tilde{\delta}$,
we continue the region, where Eq. (\ref{22}) is valid, to $\rho=+\infty$.
According to (\ref{25}), we expand the functions $\chi_{0}(\rho)$ (with $\chi_{+}(r)=0$) and  $f(\rho)$
in functions (\ref{28}) and pass to the dimensionless functions
$\psi_{2j+1}(\rho)=\sqrt{a_{ho}}\Psi_{2j+1}(r)$. Next let us multiply (\ref{22})
by $4\pi \psi_{2k+1}(\rho)$ and integrate over $\rho$ from $0$ to $\infty$.
We use the orthogonality of eigenfunctions:
 \begin{equation}
\int\limits_{0}^{\infty}4\pi\psi_{2k+1}(\rho)\psi_{2j+1}(\rho)d\rho=\delta_{k,j},
     \label{32} \end{equation}
where $\delta_{k,j} $ is the Kronecker delta.
In the integrals, we change $\ddot{\psi}_{2j+1} \rightarrow (\rho^{2}-4j-3)\psi_{2j+1}$ according to (\ref{26}) and (\ref{29}).
After such a procedure, relation (\ref{22}) is reduce to the equation
 \begin{equation}
b_{2k+1}(3I_{k,k}-4k-3-2p^{2})+d_{2k+1}\varepsilon+\sum\limits_{j\neq k}3b_{2j+1}I_{k,j}=g_{2k+1},
     \label{33} \end{equation}
where $k, j=0,1,2,\ldots, \infty$, and
 \begin{equation}
g_{2k+1}=d_{2k+1}(4k+3)-\sum\limits_{j=0}^{\infty}d_{2j+1}I_{k,j},
     \label{34} \end{equation}
\begin{equation}
I_{k,j}=4\pi\int\limits_{0}^{\infty}\rho^{2}\psi_{2k+1}(\rho)\psi_{2j+1}(\rho)d\rho,
     \label{35} \end{equation}
\begin{eqnarray}
&d_{2j+1}&=4\pi\int\limits_{0}^{\infty}\chi_{0}(r)\Psi_{2j+1}(r)d r   \label{36}  \\
&&\approx \sqrt{\frac{30\pi}{p^{3}}}\int\limits_{0}^{p-\tilde{\delta}}\rho
\psi_{2j+1}(\rho)\sqrt{1-\frac{\rho^{2}}{p^{2}}}d\rho \quad (\tilde{\delta}=\delta/a_{ho}).
   \nonumber  \end{eqnarray}

After the expansion of the functions $\chi_{0}(\rho)$ and $f(\rho)$
according to (\ref{25}) with odd $n$, Eq. (\ref{24}) takes the form
 \begin{equation}
\sum\limits_{j=0}^{\infty}b_{2j+1}d_{2j+1}=0.
     \label{37} \end{equation}
In (\ref{33}) and (\ref{37}), we change $b_{2j+1}\rightarrow c_{j+1}$ and $b_{2k+1}\rightarrow c_{k+1}$ and
then replace $j\rightarrow j-1$ and $k\rightarrow k-1$. Relations (\ref{33}) and (\ref{37})
pass to the final equations
\begin{eqnarray}
&&c_{k}(-4k+1-2p^{2})+d_{2k-1}c_{J_{max}} \nonumber \\
&&+\sum\limits_{j=1 }^{J_{max}-1}3c_{j}I_{k-1,j-1}=g_{2k-1},
     \label{38} \end{eqnarray}
 \begin{equation}
\sum\limits_{j=1}^{J_{max}-1}c_{j}d_{2j-1}=0.
     \label{39} \end{equation}
Here, $\varepsilon$ is denoted as $c_{J_{max}}$, $k$ takes the values $k=1,2,\ldots, J_{max}-1$, and the summation is cut
on some finite  $J_{max}-1\gg 1$ (instead of infinity).
\begin{figure}[ht]
\centerline{\includegraphics[width=85mm]{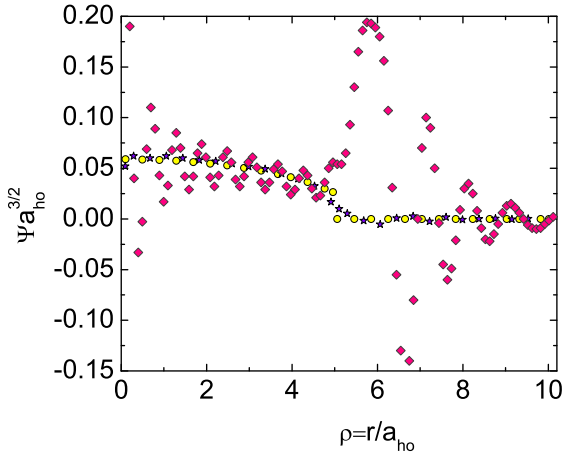} } \caption{
Solution $\Psi(\rho)$ for $\tilde{\delta}=0.5$ and $J_{max}=30$ in the
Thomas--Fermi approximation (\ref{6}), (\ref{12}) with $\chi_{+}(r)=0$, $p=5.55753$ (circles)
and with regard for the correction $f(\rho)$ (\ref{11}): for $p=5.55753$  (near the critical point; diamonds)
 and for $p=5.41$ (far from this critical point, in the middle between it and the next
 critical point; stars).
 \label{fig1}}
\end{figure}

Equations (\ref{38}) and (\ref{39}) set the
inhomogeneous system of $J_{max}$ linear equations for $J_{max}$ unknown $c_{k}$.
We solved this system numerically for various $p$, $\tilde{\delta},$ and $J_{max}$. It turns out that,
for $p \gsim 4,$    $\tilde{\delta}\ll p$ and far from the critical points $p_{cp}$,
the corrections $f(\textbf{r})$ and $\delta E$  are small in modulus as compared with
$\chi_{0}(\textbf{r})$ and $E^{s}_{0}=p^{2}\hbar\omega/2,$ respectively, and
depend slightly on $p$ and  $\tilde{\delta}$ (see Fig. 1). This indicate that
 the Thomas--Fermi approximation (\ref{6}) is close to the exact solution, as was assumed by us and was found earlier
in Ref.~\onlinecite{edwards1995,edwards1996,stringari1996,hau1998,pit1999}.
For $p\gsim 4$ and far from $p_{cp}$, our solution $\Psi(\rho)$ is close to the numerical solutions
\cite{edwards1995,stringari1996} obtained by other methods.

As $\tilde{\delta}\rightarrow 0,$ the corrections $f(\textbf{r})$ and $\delta E$ increase,
which is related to the divergence of the first and second derivatives
of $\Psi^{s}_{0}(r)$ (\ref{6}) as $r\rightarrow R$.

\begin{figure}[ht]
\centerline{\includegraphics[width=85mm]{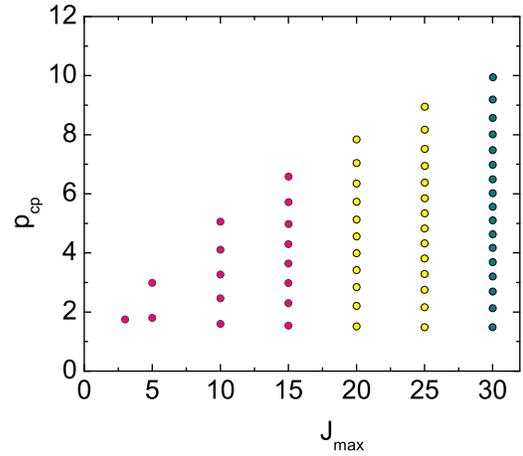} } \caption{
The critical points $p_{cp}$, at which the determinant of the system of equations
(\ref{38}), (\ref{39}) turns into zero, for various $J_{max}$. \label{fig2} }
\end{figure}

The most interesting result consists in the discovery of the
critical points (Fig. 2) that are values of the parameter $p $, at which
the determinant of matrix (\ref{38}), (\ref{39}) turns to zero.
This implies that one or several
coefficients $c_{k}$ (in the collection of solutions $\{ c_{k}\}$)
are arbitrary and can be arbitrarily large. Since the solutions
$c_{k}$ are inversely proportional to the matrix determinant, they
increase in modulus as $p $ approaches one of
the critical points. Therefore, the corrections $f(r)$ (\ref{25})
and $\delta E$ increase as well.  If $p $ is very close to the
critical point $p_{cp}$, then $|f(r)|$ and $|\delta E|$ become larger
than $\chi_{0}(r)$ and $E^{s}_{0}$. This means that, in a small vicinity of $p_{cp}$, the exact
solution must strongly differ from the Thomas--Fermi approximation (\ref{6}). Such values of $p$ form
the critical region. This is illustrated in Fig. 1, where the
stars and the rhombs show the solution $\Psi(\rho)$ far from
and near the chosen critical point $p_{cp},$ respectively. For the
rhombs, the value of $p$ is such that $|det|$ of matrix
(\ref{38}), (\ref{39}) by 100 times less than $|det|$ for the
``background'' $p$ corresponding to the curve with stars. In
addition, the corrections for the curve with rhombs are large and
such that $|f(\rho)| \sim \chi_{0}(\rho)$ in the region  $\rho <
p-\tilde{\delta}$; therefore, the value of $p$ for the curve with
rhombs determines the half-width $\lambda^{cp}$ (see below).

The values of $p_{cp},$ which are larger than 1, are presented in Fig. 2 for various $J_{max}$.
It is seen that the number of critical points $p_{cp}$ increases with $J_{max}$.
In this case, the new $p_{cp}$ arise from above, so that the net of values of $p_{cp}$ becomes denser in the region with large $p$.
For $J_{max}=\infty,$ the number of critical points $p_{cp}$ should be, apparently, infinite. The new (as compared with Fig. 2)
points should be in the region with large $p$ and should come to infinity. We arrived only at $J_{max}=30$.
Further the numerical analysis gives distorted values due to, probably, the appearance
of too large numbers ($> 10^{100}$) in (\ref{30}), with which the computer program cannot work.
With regard for the dynamics of points already obtained, we expect that, 1) in the region $p< 4$,
the exact solution for $J_{max}=\infty$ will give
the values of $p_{cp}$ insignificantly differing from those obtained for $J_{max}=30$; 2)
the net of $p_{cp}$ will become denser in the region $4<p< 10$; and 3)
the infinite number of new $p_{cp}$ will appear in the region $p >10.$
It is seen from Fig. 2 that the largest $p_{cp}$ depends on $J_{max}$.
For the given $J_{max},$ the greatest number of the considered basis function
$\Psi_{n}(r)$ (\ref{28}) is $n_{max}= 2(J_{max}-2)+1$.
The value of the largest $p_{cp}$ is determined by the largest
$r$, for which $\Psi_{n_{max}}(r)$ is not small: $p_{cp}\simeq r_{max}(n_{max})/a_{ho}$.
As $J_{max}\rightarrow \infty,$ we obtain $n_{max}\rightarrow \infty$,
$r_{max}\rightarrow \infty,$ and, therefore, $p^{max}_{cp}\rightarrow \infty$.

\begin{figure}[ht]
\centerline{\includegraphics[width=85mm]{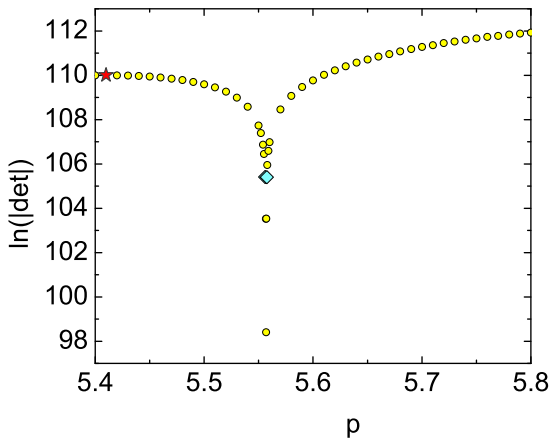} } \caption{
Circles mark the values of the logarithm of determinant modulus of
matrix (\ref{38}), (\ref{39}) in a vicinity of the critical point
$p_{cp}\approx 5.55688$ for $\tilde{\delta}=0.5$ and $J_{max}=30$. Two
close rhombs (merging in a single one) indicate two points
determining the line width. The center of the line is located
in the middle between these points. The profile of $\Psi(\rho)$
for the value of $p$, corresponding to one of these rhombs,  is
shown by rhombs in Fig. 1. The star indicates the value of $p$
corresponding to the curve with stars in Fig. 1.
 \label{fig3}}
\end{figure}
In Fig. 3, we show the values of the determinant in a vicinity of
the critical point. Note that the
determinant is positive to the
left from the minimum $p=p_{cp}$ and is negative to the right. Due
to the narrowness of lines, one can find the critical points by the
change of a sign of the determinant.

It is convenient to introduce a critical half-width $\lambda^{cp}$ equal
to the modulus of the difference between the value of $p$,
at which $det=0$, and the value of $p$, at which the correction
$|f(r)|$ is approximately equal to the bare one $\chi_{0}(r)$ for all $\rho < p-\tilde{\delta}$.
For $J_{max}=30$ and $\tilde{\delta}=0.5,$ we obtained $\lambda^{cp}\simeq 6.5\cdot 10^{-4}$ for the critical point
$p_{cp}\approx 5.557.$ Whereas, for
$p_{cp}\approx 9.19,$ we have $\lambda^{cp}\simeq 4.5 \cdot 10^{-4}$.
That is, $\lambda^{cp}$ changes slightly with increasing $p_{cp}$.

For the experimental discovery of a critical point,
it is necessary to change the number of atoms $N$ or the frequency (magnetic field) $\omega$
\textit{very smoothly}, with the step
 \begin{equation}
\frac{{\sss \triangle} \omega}{\omega}=\frac{2{\sss \triangle} N}{N}=
\frac{10{\sss \triangle}p}{p}=\frac{\lambda^{cp}}{p}\sim  10^{-4}.
     \label{40} \end{equation}
 Here, we took relation (\ref{8}) into account
and chose the step ${\sss \triangle}p=\lambda^{cp}/10$.
The step should be at least several times less than $\lambda^{cp}$. Since $\lambda^{cp}$ was estimated only approximately,
it is better to choose the step to be smaller (in order not to miss the line).
Therefore, we chose $\lambda^{cp}/10$.

\section{Discussion}
The significant point is the approximation in use:
instead of the exact equation  (\ref{13}) taking the tail  $\chi_{+}(\rho)$ into account, we solved
the approximate equation (\ref{20}), by extending it from the interval $[0,p-\tilde{\delta}]$
onto the whole semiaxis $[0,+\infty]$. These equations differ in the region $\rho > p-\tilde{\delta}$. The use of
(\ref{13}) instead of (\ref{20}) will lead to a change of coefficients in
matrix (\ref{38}). This shifts the critical points. The shift will be, most likely, small,
since the functions are small at $\rho > p-\tilde{\delta}$.
But, in principle, the critical points can disappear entirely.
For  $\tilde{\delta}=0.001, 0.1,$ and $0.5$ in formula (\ref{36}),
the values of $p_{cp}$ differ by $0.01$ on the average,
i.e., less than by $0.5\%$ (for $J_{max}=25$ and $30$).
This is an argument in favor of that the shift should be small.

Note that  we have
found the critical points within analogous approximations also for the one-dimensional problem.
Now, we study the case of a cylindrical trap.

The time-dependent GP equation was considered \cite{ruprecht1995} and it was found that the solution
for the ground state is stable for two tens of values of the parameter $C_{nl}= N\cdot const$
and that its energy depends smoothly on $C_{nl}$.
This does not contradict our results, since the critical regions are very narrow,
and it is necessary to take $\sim 10^3$ values of $C_{nl}$ in order to accidently fall in such a
region. We believe that
the solution exists in the critical regions, but it has the different energy as compared with adjacent noncritical points.
In this case, the ground-state energy must have a spike in the critical region.

Let the critical points exist in the case where the tail of the condensate wave function (WF) is considered. Which is the solution for
$p=p_{cp}$? Two versions are possible: 1) the solution differs quantitatively from the Thomas--Fermi approximation (\ref{6}),
but it is qualitatively similar to it and has no nodes; 2) the solution differs from (\ref{6}) even qualitatively and has nodes
and a very nonclassical shape. The second version is more interesting, but there is some limitation for it. The total WF
$\Psi(\textbf{r}_{1},\ldots,\textbf{r}_{N_{f}},t)$ describing both the condensate and noncondensate atoms
satisfies the linear Schr\"{o}dinger equation and, therefore, must have no nodes in the ground state.  If we write approximately the total WF as
\begin{equation}
\Psi(\textbf{r}_{1},\ldots,\textbf{r}_{N_{f}},t)\approx \prod\limits_{j=1}^{N_{f}}\Psi_{c}(\textbf{r}_{j},t),
     \label{41} \end{equation}
then the Schr\"{o}dinger equation yields the nonstationary GP equation for $\Psi_{c}(\textbf{r},t)$ \cite{gross1963}. Relation (\ref{41}) assumes that
\textit{all} atoms belong to the condensate, which is wrong.
The condensate WF $\Psi_{c}(\textbf{r},t)$   satisfies the GP equation.
The well-known quantum-mechanical theorem (Ref.~\onlinecite{gilbert}, Chap. 6)
is inapplicable to GP equation due to its nonlinearity, so that the ground-state WF of the condensate may have nodes.
However, the total ground-state WF
has no nodes, and it is unclear whether this fact is consistent with the presence of nodes of the condensate WF.
If not, then the condensate WF must have no nodes for $p=p_{cp}$.

The GP equation was comprehensively analyzed in Ref.~\onlinecite{lieb2000}, where it was asserted, in particular,
that the ground-state WF  $\Psi_{0}(\textbf{r})$ has no nodes (Theorem 2.1). This was proved in Lemma A.4 on the following base: if the function $\Psi_{\infty}(\textbf{r})$  minimizes the functional
\begin{equation}
\varepsilon(\Psi)=N\int\limits_{R^3}\left [(\nabla \Psi(\textbf{r}))^{2}+
U(\textbf{r})|\Psi(\textbf{r})|^{2}+4\pi Na|\Psi(\textbf{r})|^{4}  \right ]d\textbf{r}
     \label{42} \end{equation}
(we wrote it for the normalization $ \int d \textbf{r} |\Psi(\textbf{r})|^{2}=1$, like the GP equation (\ref{1})),
then the function $|\Psi_{\infty}(\textbf{r})|$ is
also a minimizing one in view of $\varepsilon(|\Psi|)\leq \varepsilon(\Psi).$
Therefore, the nonnegative  $|\Psi_{\infty}|$ must describe the ground state.
This reasoning seems to us not quite strict. Assume the contrary: let the ground state be described by a
wave function $\Psi_{0}(\textbf{r})=\Psi_{0}(r)$ equal to zero at  $r=r_{1}$.
It is obvious that the functional
$\varepsilon(|\Psi_{0}|)$ differs from the functional $\varepsilon(\Psi_{0})$ only due to the
first term in (\ref{42}), since the derivative $\nabla_{r} \Psi_{0}(r)$ is continuous at the point $r=r_{1}$,
but $\nabla_{r} |\Psi_{0}(r)|$ changes by jump due to the change of a sign of $\Psi_{0}(r)$ at the point $r=r_{1}$.
However, the singularity is present only at this single point. At all remaining points, the derivatives are the same in modulus. Therefore,
$\varepsilon(|\Psi_{0}|)= \varepsilon(\Psi_{0})$.  In this case,  $\partial^{2}|\Psi_{0}(r)|/\partial r^2 =\infty$
at the point $r=r_{1}$, whereas the remaining terms in the GP equation (\ref{1}) are finite. That is 
the function $|\Psi_{0}|$ is not a solution of the GP equation. Thus,
the reasoning in Ref.~\onlinecite{lieb2000} does not refute our assumption. In other words, $\Psi_{0}(\textbf{r})$
\textit{can have nodes} in principle. But it can have no nodes as well. The question about nodes of the ground-state WF for the nonlinear GP equation
should be separately studied.

We address the unsolved questions to the future. \\

\section{Conclusion}
The main result consists in the discovery of narrow critical regions, composed of values of the parameter
$p=R/a_{ho}=(15 Na/a_{ho})^{1/5}$, at which the solution for the ground-state wave function
of the condensate differs strongly from
the Thomas--Fermi approximation. The result is valid at the neglect of the tail of the wave function.
Such specific features were not found earlier. Of course, it is important to verify the solution. For that,
it is necessary to find a solution with regard for the WF tail and to make sure of the presence of critical points.
It is of interest to go over a wide band of values of
$p$ experimentally with the step ${\sss \triangle} p \ll \lambda^{cp}$.
As $p$ approaches the critical points $p_{cp},$ the cloud of the condensate must strongly change the size.
It would be especially interesting if the cloud would take a very nonclassical shape 
in a close vicinity of $p_{cp}$.
Such an effect would be one more clear manifestation of quantum laws on macroscopic scales.

 \renewcommand\refname{}



 \end{document}